\documentstyle[epsfig,sprocl]{article}

\def\NPB{{\em Nucl. Phys.} B}
\def\PLB{{\em Phys. Lett.}  B}
\def\PRL{{\em Phys. Rev. Lett.} }
\def\PRD{{\em Phys. Rev.} D}

\def\be{\begin{equation}}
\def\ee{\end{equation}}
\def\bea{\begin{eqnarray}}
\def\eea{\end{eqnarray}}

\def\beq{\begin{equation}}
\def\eeq{\end{equation}}
\def\bea{\begin{eqnarray}}
\def\eea{\end{eqnarray}}
\def\bem{\begin{math}}
\def\eem{\end{math}}
\def\bit{\begin{itemize}}
\def\eit{\end{itemize}}
\def\bla{\begin{flushright}}
\def\ela{\end{flushright}}
\def\qq2{$Q^2$}               
\def\aa1{$A_1(x,Q^2)$}        
\def\ff1{$F_1(x,Q^2)$}        
\def\gg1{$g_1(x,Q^2)$}        

\begin{document}

\title{ABOUT THE $Q^2$ DEPENDENCE OF THE MEASURED ASYMMETRY $A_1(x)$}

\author{A. V. KOTIKOV, D. V. PESHEKHONOV}

\address{Particle Physics Laboratory, JINR \\ Dubna, 141980 Russia \\
E-mail: kotikov@sunse.jinr.ru, peshehon@sunse.jinr.ru}


\maketitle\abstracts{ 
We propose the new approach for taking into account the $Q^2$ dependence of
measured asymmetry $A_1$.
This approach is based on the similarity of the $Q^2$ behaviour 
and the shape of 
the spin-dependent structure function $g_1(x,Q^2)$ and spin averaged 
structure function $F_3(x,Q^2)$. 
The analysis is applied on 
 the SMC and E154
experimental data.}



An experimental study of the nucleon spin structure is realized by
measuring of the asymmetry $A_1(x,Q^2) = g_1(x,Q^2) / F_1(x,Q^2)$.
The most known theoretical predictions on spin dependent structure
function $g_1(x,Q^2)$ of the nucleon were done by Bjorken \cite{Bj} and
Ellis and Jaffe \cite{EJ} for the so called {\it first moment value}
$\Gamma_1 = \int_0^1 g_1(x) dx$.\\
Studying the properties of $g_1(x,Q^2)$ and 
the calculation of the $\Gamma_1$ value require the knowledge of
structure function $g_1$ at the same $Q^2$ in the hole $x$ range.
Experimentally asymmetry $A_1$ is measuring at different values of $Q^2$
for different $x$ bins.
An accuracy of the
modern experiments
\cite{EG,SMC,E154n,Q2E154}
allows to analyze data in the assumption \cite{EK93}
that asymmetry \aa1 is \qq2
independent (structure functions $g_1$ and $F_1$ have the same $Q^2$
dependence)
\begin{eqnarray}
A_1(x,Q^2) = A_1(x) \label{a1}
\end{eqnarray}
But the precise checking of the Bjorken and Ellis - Jaffe sum
rules requires considering the $Q^2$ dependence of $A_1$ or $g_1$.
Moreover, the assumption (1) the asymmetry \aa1 is 
$Q^2$ independent 
does not  follow from the theory. 
On the contrary, the 
behaviour of $F_1$ and $g_1$ as a functions of $Q^2$ is expected to be 
different due to the difference between polarized and unpolarized splitting 
functions.\\
%
There are several approaches (see \cite{Q2A,GRSV} and its references) 
 how to take into account 
the $Q^2$ dependence of $A_1$. They are
based on different approximate solutions of the DGLAP equations.
Some of them have been used already by Spin Muon Collaboration (SMC)
and E154 Collaboration in the last analyses of experimental data
(see \cite{SMC} and \cite{Q2E154}, respectively). \\

In this article we suggest to use another idea which is based on the 
observation that the splitting functions of  the DGLAP equations for
the SF $g_1$ and  $F_3$ and  the shapes of the SF themselves
are the similar in a wide 
$x$ range and, thus, the $Q^2$ 
dependence of them has to be close as a consequence. 
Our approach for $Q^2$-dependence of $A_1$ are very simple 
(see eq.(\ref{5})) and leads to the results, which are very similar 
to ones based on the  DGLAP evolution.\\
To demonstrate the validity of the observation, we note that
the r.h.s. of DGLAP equations for
NS parts of $g_1$ and $F_3$ is the same
(at least in first two orders of the perturbative QCD)
and differs from $F_1$ already in the first subleading order.
For the singlet part of $g_1$ and for $F_3$ the difference between
perturbatively calculated spliting functions
is also negligible
(see \cite{KP,KNP}).
This observation allows us to conclude the function 
\bea
A_1^*(x) = {g_1(x,Q^2) \over F_3(x,Q^2)} \nonumber
\eea
should be practically $Q^2$ independent 
and
the asymmetry $A_1$ at some  $Q^2$ can be defined than as :
\bea
A_1(x_i,Q^2) =  {F_3(x_i,Q^2) \over F_3(x_i,Q^2_i)} \cdot
{F_1(x_i,Q^2_i) \over F_1(x_i,Q^2)} \cdot A_1(x_i,Q^2_i),
\label{5}
\eea
where $x_i$ ($Q^2_i$) means an experimentally measured value of $x$ ($Q^2$).\\

To apply the proposed approach we use the 
SMC \cite{SMC} and  E154 Collaboration \cite{E154n,Q2E154}
data.
To use relation (\ref{5}) we parametrize CCFR data on 
$F_2(x,Q^2)$ and $xF_3(x,Q^2)$ 
\cite{CCFRN} in the same form as NMC fit of the structure function 
$F_2(x,Q^2)$ \cite{NMC} (see \cite{KNP}). 
To obtain structure function $F_1(x,Q^2)$ we also take the parametrization of 
the CCFR data on $F_2(x,Q^2)$ \cite{CCFRN} and the SLAC parametrization of 
$R(x,Q^2)$ \cite{SLAC} 
and use relation :
\begin{equation}
F_1(x,Q^2)= \frac{F_2(x,Q^2)}{2x(1+R(x,Q^2))} \cdot 
(1+ \frac{4M^2x^2}{Q^2}) ,
\label{5.1}
\end{equation}
We use in Eq.(\ref{5}) parametrizations of CCFR data \cite{CCFRN} for both 
SF $xF_3(x,Q^2)$ and $F_2(x,Q^2)$ to avoid systematical uncertanties and
nucleon correlation in nuclei.\\
 \hskip -.56cm
The SF $g_1(x,Q^2)$ is calculated using the asymmetry $A_1(Q^2)$ as
\bea
g_1(x,Q^2)= A_1(x,Q^2)\cdot F_1(x,Q^2), 
\label{5.2}
\eea 
where spin average SF $F_1$ has been calculated using NMC 
parametrization of $F_2(x,Q^2)$ \cite{NMC}.
The results are presented in Fig. 1 and Fig. 2 for E154 and SMC data,
respectively. Our results are in an excelent agreement with the calculations
which are based on direct DGLAP evolution.
\vskip 2.7cm \hskip 1.5cm
\begin{tabular}{c}
\begin{minipage}[h]{5.cm}
\epsfig{figure=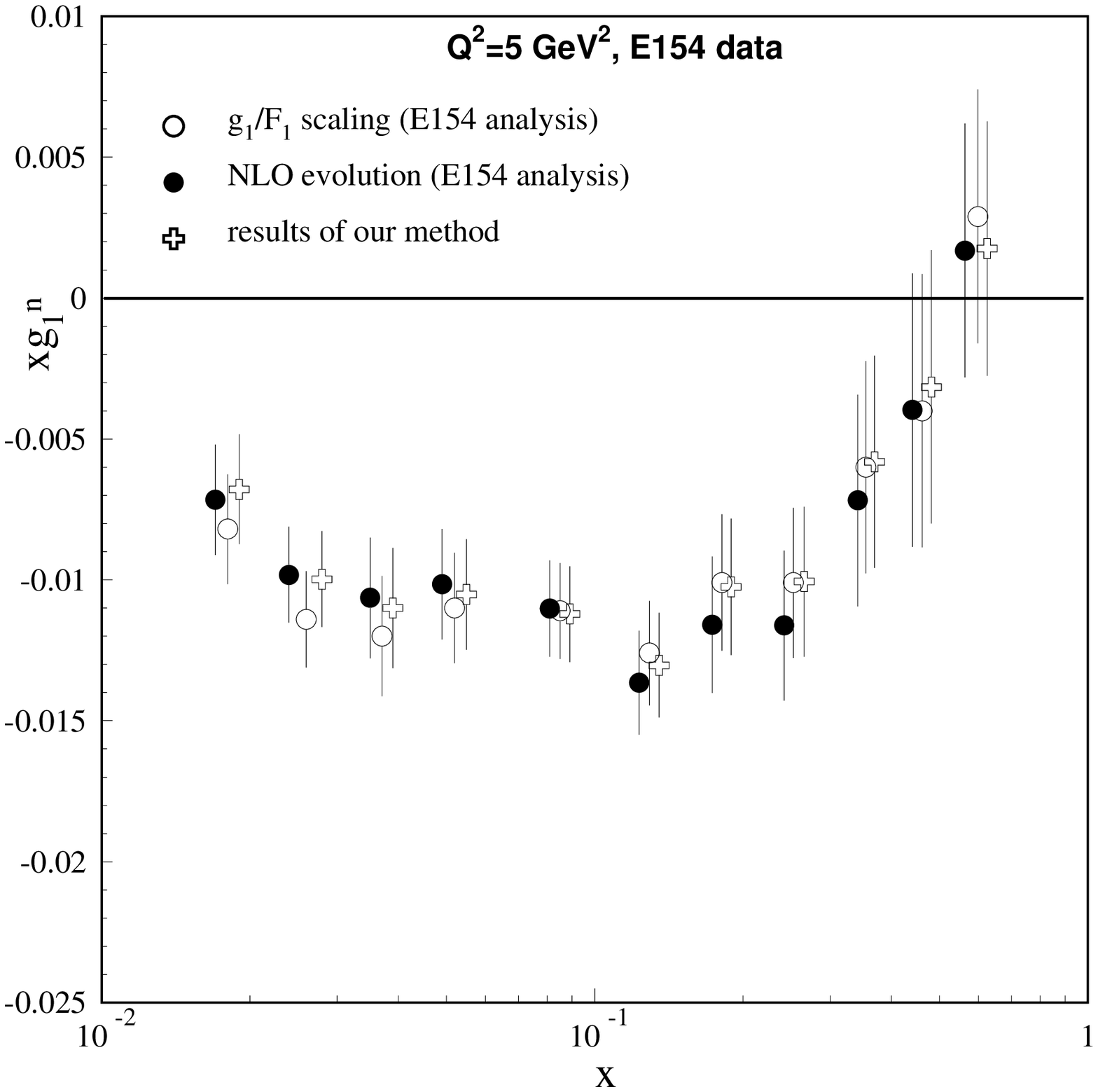,width=5.cm}
\end{minipage}
\end{tabular}
\vskip -1.2cm
\hskip -.56cm
{\bf Figure 1.}The structure function $xg_1^n(x,Q^2)$ evolved to $Q^2=5GeV^2$
using our eq.(2), DGLAP NLO evolution and the assumption that $g_1^n/F_1^n$
is $Q^2$ independent. Last two sets are from$^6$. 
\vskip 2.2cm \hskip 1.9cm
\begin{tabular}{c}
\begin{minipage}[h]{5.cm}
\epsfig{figure=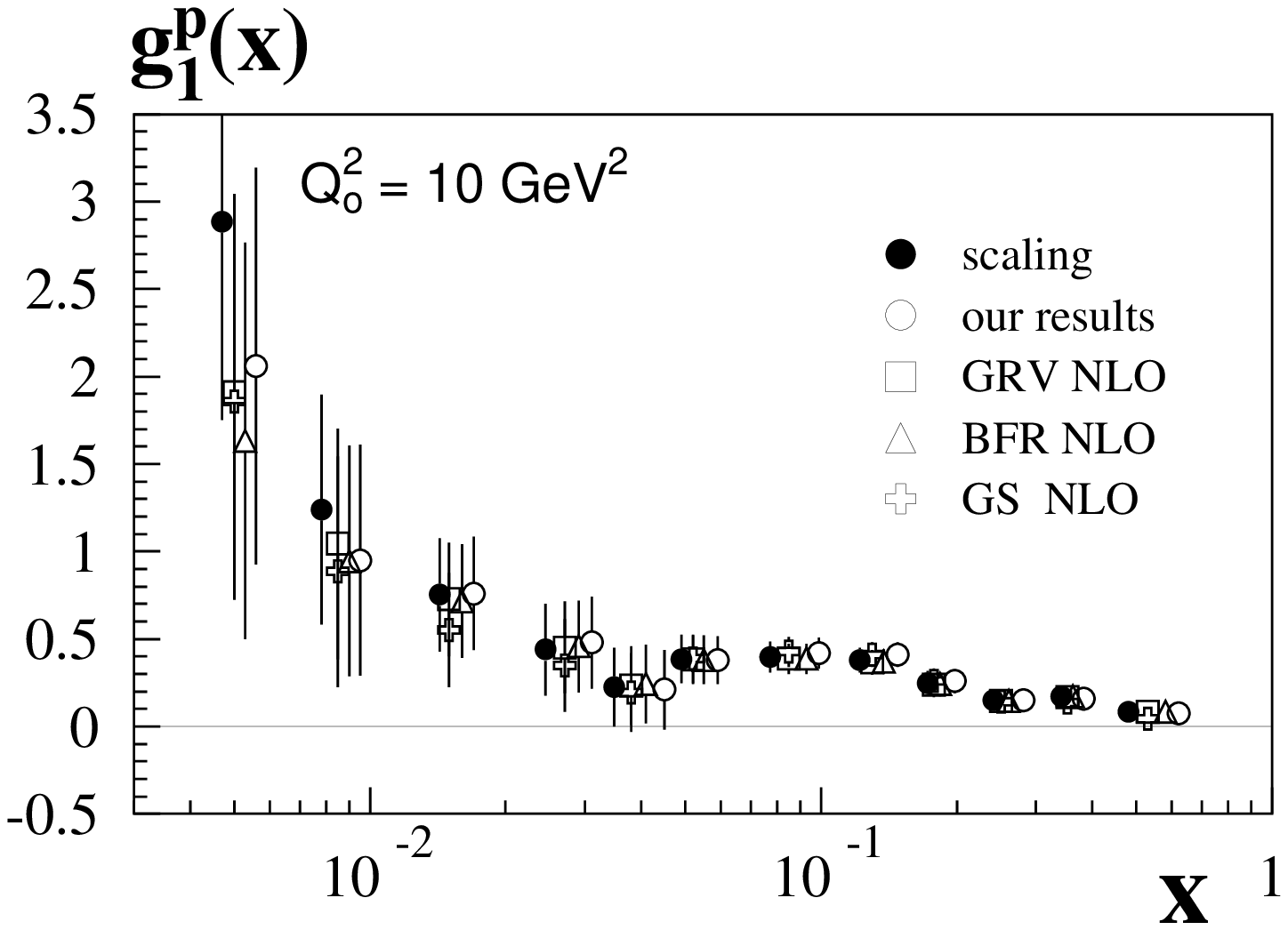,width=5.cm}
\end{minipage}
\end{tabular}
\vskip -1.9cm
\hskip -.56cm
{\bf Figure 2.} The structure function $xg_1^n(x,Q^2)$ evolved to $Q^2=10GeV^2$
using our eq.(2), the assumption that $g_1^n/F_1^n$
is $Q^2$ independent, and DGLAP NLO evolution according to the analysis$^9$.\\
\vskip -.2cm \hskip -.56cm
To make another comparison with the theory we 
have calculated also the first moment
value of the structure function $g_1$ at different $Q^2$.
Using eq.(\ref{5}), we recalculate the SMC measured asymmetry of the
proton and deuteron and E154 one of neutron
 at $Q^2= 100~ {\rm GeV}^2$, $30~ {\rm GeV}^2$,
$Q^2= 10~ {\rm GeV}^2$ and $3~ {\rm GeV}^2$
and get the value of $\int g_1(x) dx$ through the
measured $x$ ranges.  To obtain the first moment values
$\Gamma_1^{p(d)}$ we use an original estimations of SMC and E154
for unmeasured regions.
As the last step we calculate the difference $\Gamma_1^p - \Gamma_1^n$
(for SMC proton and deutron data  $\Gamma_1^p - \Gamma_1^n =
2 \Gamma_1^p -
2 \Gamma_1^d / (1-1.5 \cdot \omega_D)$ where
$\omega_D=0.05$).
In Table 1 we present the 
results for the mean values of $\Gamma_1^p - \Gamma_1^n $,
because the errors coincide with the errors of original analyses
\cite{SMC,E154n,Q2E154}\footnote{The theoretical
predictions computed in \cite{LV} to the third order in the QCD $\alpha_s$.}
The value of $\Gamma_1^p - \Gamma_1^n $ at $Q^2 = 10 Gev^2$ obtained by 
direct DGLAP evolution are taken from article \cite{SMC}.

\begin{table}[t]
\caption{The mean values of $\Gamma_1^p - \Gamma_1^n $.\label{tab:exp}}
\vspace{0.2cm}
\begin{center}
\footnotesize
\begin{tabular}{|l|c|c|c|c|}
\hline
$Q^2$ (GeV$^2$) & 100 & 30 & 10 & 3 \\
\hline
\multicolumn{5}{|c|}{
SMC proton and deutron data}\\
\hline
$A_1$-scaling & 0.297 & 0.226 & 0.202 & 0.170 \\
\hline 
Evolution &  &  & 0.183 &  \\
\hline 
$A_1^*$-scaling & 0.210 & 0.201 & 0.191 & 0.176 \\
\hline
\multicolumn{5}{|c|}{
SMC proton and E154 neutron data}\\
\hline
$A_1$-scaling & 0.221 & 0.209 & 0.194 & 0.170 \\
\hline 
$A_1^*$-scaling & 0.194 & 0.190 & 0.185 & 0.175 \\
\hline
Theory & 0.194 & 0.191 & 0.187 & 0.180 \\
\hline
\end{tabular}
\end{center}
\end{table}

\vspace{0.5cm}

Let us now present the main results, which are following from the Table 1 
and the  Figures.
\begin{itemize}
%
\item
The results are in excelent agreement with $g_1(x,Q^2)$ data of SMC and
E154 Collaborations, based 
on direct DGLAP evolution.
\item Our method allows to test of the Bjorken sum rule in a simple way 
with a good accuracy. 
Obtained results on the $\Gamma_1^p - \Gamma_1^n$ show that used experimental 
data well confirm the Bjorken sum rule  prediction.
\end{itemize}
\hskip -.56cm

\section*{Acknowledgments}

One of us (AVK) is gratefull very much to Organizing Committee of 
Workshop DIS98
for the financial suppor. 

\newpage


\end{document}